\documentclass{article}
\input amssym.def
\input amssym.tex
\usepackage[a4paper]{geometry}

\def\ben{\begin{equation}}
\def\een{\end{equation}}
\def\half{\frac{1}{2}}
\def\bea{\begin{eqnarray}}
\def\eea{\end{eqnarray}}

\def\p{\partial}
\def\mathbb{\Bbb}

\def\ben{\begin{equation}}
\def\een{\end{equation}}
\def\half{{1 \over 2}}
\def\bea{\begin{eqnarray}}
\def\eea{\end{eqnarray}}

\def \p{\partial}

\def\p{\partial}

\newcount\hour \newcount\minute
\hour=\time  \divide \hour by 60
\minute=\time
\loop \ifnum \minute > 59 \advance \minute by -60 \repeat
\def\nowtwelve{\ifnum \hour<13 \number\hour:%           % supresses leading 0's
                      \ifnum \minute<10 0\fi%           % so add it it
                      \number\minute
                      \ifnum \hour<12 \ A.M.\else \ P.M.\fi
         \else \advance \hour by -12 \number\hour:%     % supresses leading 0's
                      \ifnum \minute<10 0\fi%           % add it in
                      \number\minute \ P.M.\fi}
\def\nowtwentyfour{\ifnum \hour<10 0\fi%                % need a leading 0
                \number\hour:%                          % supresses leading 0's
                \ifnum \minute<10 0\fi%                 % add it in
                \number\minute}

\begin{document}
\title{
\begin{flushright}
 \small{UPR-1236-T} \\ 
\small{DAMTP-2012-11}
\end{flushright}
\vspace{0.2in}
{\Large Graphene and the Zermelo Optical Metric of the BTZ Black Hole
}}
\author{\large M. Cveti\v c$^{1,2}$ and G.W Gibbons $^{3}$  
\\
\\   \small{$^1${\it Department of Physics and Astronomy,
University of Pennsylvania, Philadelphia, PA 19104-6396, USA} }
\\ \small{$^2${\it  Center for Applied Mathematics and Theoretical Physics,
University of Maribor, Maribor, Slovenia}}
\\   \small{$^3${\it  D.A.M.T.P., University of  Cambridge, 
Wilberforce Road, Cambridge CB3 0WA, U.K. }              }  
\\ }
\maketitle
\begin{abstract}
It is well known  that the low energy electron excitations  of  the curved graphene sheet $\Sigma$    are  solutions of the massless Dirac equation on a $2+1$ dimensional ultra-static metric on ${\Bbb R} \times \Sigma$.  An externally applied electric field  on the graphene sheet induces a gauge potential which  could be mimicked by considering a stationary optical metric of the  Zermelo  
form, which  is conformal to the BTZ black hole when the sheet has a constant 
negative curvature. The Randers form of the metric can model a magnetic field, 
which is related  by a boost to an  electric  one  in the Zermelo frame.  We also show that  there is fundamental geometric obstacle to obtaining a model that extends all the  way to the black hole horizon.

\end{abstract}

\eject
\pagebreak
%\tableofcontents
\newpage

\section{Introduction} 

In a recent paper \cite{Iorio} it has been suggested  that a 
  surface  of revolution with constant negative
curvature ( Beltrami Trumpet) made from graphene
may exhibit  some of the interesting effects 
which arise in  quantum field theory in a curved spacetimes.

Of particular interest in this connection 
would be effects associated with  
BTZ black hole   metrics \cite{BTZ}. These are solutions of
Einstein's equations in $2+1$ dimensions with negative cosmological constant
and have been intensively studied
in recent years because  they also arise in string theory descriptions
of black holes and their quantum microstates.    The metric of the BTZ black hole is specified by the mass $M$, angular momentum $J$ and the negative cosmological constant $\Lambda \equiv -l^{-2}$.  Its metric is of the form:
\ben
ds^2  _{BTZ} = -\Delta \ dt ^2 + \frac{dr^2}{\Delta} + r^2
 \big( d \phi  -\frac{J}{2 r^2} dt  \big)^2\, , 
  \label{BTZm}
  \een
  where
  \ben
\Delta(r)= \frac{ r^2}{l^2}-M + \frac{J^2}{4 r^2}.  \label{cf}
\een

It is well known (see \cite{Cortijo} for a review)
that  on a  curved graphene sheet $\Sigma$  
there are low energy electronic
excitations  which  satisfy the  massless  Dirac equation on 
${\Bbb R}\times \Sigma$
with respect to the metric
\ben
ds ^2 = -dt ^2 + h_{ij}dx^idx^j  \,,\qquad i=1,2 \label{ultrastatic} 
\een 
where $h_{ij}$ is the metric induced on the sheet $\Sigma$. 
The case considered in \cite{Iorio} is  a surface of revolution  in
Euclidean three space ${\Bbb E} ^3$
with coordinates $(x(x^i) ,y(x^i) ,z(x^i) )$.

The metric  (\ref{ultrastatic}) is an example of an ultra-static metric.
That is,  it is static, i.e.   invariant under both time translations
and time reversal.  The general static metric takes the local  form
 \ben
 ds ^2 = -V dt ^2 + g_{ij} dx^i dx^j \,.
\een
For a static black hole  we have $V>0$ outside the horizon
which is located at $V=0$.  A static metric  is ultrastatic if $V=1$.  
An ultra
static metric is one for 
which the  gravitational red shifting does not occur, i.e
one for which $g_{tt}$ is independent of the spatial coordinates.
As a consequence a massive  particle may remain at rest
in such a spacetime because  it suffers  no gravitational attraction.
Since in the case of graphene,  there is no obvious source
of  red shifting, the  assumption made in \cite{Iorio} 
that $g_{tt}=-1$ appears to be 
physically well justified. 
Clearly a black hole metric cannot be ultra-static.

However the massless Dirac equation 
is invariant under conformal rescalings, that is, in 
$D$ spacetime dimensions,
if 
\ben
\tilde g_{\mu \nu}= \Omega ^2 g_{\mu \nu} \,,\qquad 
\tilde \Psi = \frac{1}{\Omega ^{\frac{D-1}{2} } }  \Psi \, , \een
 we have
\ben
 \gamma ^\mu \nabla_\mu  \Psi =  \frac{1}{\Omega ^{\frac{D+1}{2}}}
     \tilde \gamma ^\mu \tilde 
\nabla_\mu \Psi   \,.
\een 
Moreover, any static metric is locally conformally ultra static, that is 
\ben
 -V dt ^2 + g_{ij} dx^i dx^j = V \Bigr \{-dt^2 + h_{ij} dx^i dx^j \Bigl \}  
\,,\label{ultra} \een
where $h_{ij} = \frac{1}{V} g_{ij} $ is called \cite{Gibbons1} 
the {\it optical metric} 
of the static metric on the left hand side of (\ref{ultra}), {\sl  classically}
at least there might appear  to be no obstacle to mimicking the effects
of horizons  on massless fermions using 
ultra-static metrics for graphene of the form   (\ref{ultrastatic}), as 
suggested in \cite{Iorio}.
In quantum field theory however, the choice of vacuum state
is not necessarily conformally invariant and so the   mimicking 
the Unruh or Hawking effects is not obviously  possible.

However not all time-independent (i.e. stationary) metrics
are static. In particular there are rotating black holes in
in $2+1$ dimensions, including BTZ black holes, which are stationary.
Their metric take the general form
\ben
ds ^2=-V(dt+\omega_idx^i) + g_{ij} dx^i dx^j \,. 
\een

The effect of the rotation
is to introduce an so-called gravito-magnetic field, whose effect 
on particles  resembles
that of a  magnetic field.  
Now it  is possible to consider graphene 
subject to an externally applied magnetic field.
This  would induce a gauge field $A_i$ on the  graphene  sheet $\Sigma$.  
From a gravitational point of view
such a gauge connection, which breaks time reversal invariance
but not conformal invariance,
would   mimic 
a {\it stationary} rather than a static metric. 

Locally any stationary  metric
may be brought by a conformal rescaling \cite{Gibbons2} to
one of two forms \cite{Gibbons2}.  
The  {\it Randers } form   
\ben
ds_R ^2 = -(dt +\omega _i dx^i) ^2  +a_{ij} dx^i dx^j  \, , \label{Randers}
\een
where $a_{ij}= V^{-1} g_{ij} $ and   $\omega _i dx^i$ may be 
thought of as the gravito-magnetic
vector potential.  

Using a different conformal rescaling and completely the square differently
we obtain the   {\it Zermelo}  form 
\ben
ds_Z^2 = -dt^2 + h_{ij} (dx^i -W^ i dt)( dx^j -W^j dt)\, \label{Zermelo}  
\een
 where $h_{ij}$ is called the Zermelo optical metric and $W^i$ the 
{\it wind vector field}. Roughly speaking, and as we shall make more precise
later, the effect of the wind
is similar to that of an electric field. The  relation between these two
forms and their physical significance may be found in \cite{Gibbons2}.
Note that the form of the  metric we have given for the rotating BTZ
black hole (\ref{BTZm}) 
is, up to a conformal factor $\Delta$, already in Zermelo form.

The Randers optical  form of the conformally rescaled
optical metric of the BTZ black hole (\ref{BTZm})
is considerably more complicated. It may be written as 
\ben
ds_R  ^2 = -( dt + \frac{J}{2 \Delta_0} d \phi) ^2  
 + \frac{dr^2}{ \Delta \Delta_0}  + r^2 \frac{\Delta}{\Delta_0^2} d \phi^2
 \,, \een   
where
\ben
\Delta_0= \frac{r^2}{l^2}-M \,,
\een
is the conformal factor. There is an ergosphere  at the positive zero of $\Delta_o$, i.e. at 
 $r=lM$, which lies  outside the 
horizon which is located at $r=r_+$, where $r_+$ is the outermost  
zero of $\Delta$, that is 
%\ben
%r_+ =\frac{l}{\sqrt{2}}\sqrt{ M+ \sqrt{M^2 - J^2}}  
%\,.
%\een 
\ben
r_+^2=\frac{1}{2}\left(lM+\sqrt{(lM)^2-{J^2}} \right)\, . 
\een
The ergoregion leads  to complications near the
horizon, and so in what follows we shall mainly focus 
our attention on the Zermelo form.

When considering the Dirac equation, we need to introduce
an orthonormal frame or triad of vector fields 
${ e} _a$, $a=0,1,2$ . The frames $e^R_a$ and $e^Z_a$, which are natural 
 to  introduce
for the respective  Randers and the Zermelo form  of the metric, differ and  they may be thought of as 
as  boosted with respect to one another. In the Randers case the natural 
frame has a timelike leg of the form
\ben
e^R _0 = \frac{\p}{\p t} \,.  
\een    
In the Zermelo case the timelike leg  has the form
\ben
e^Z _0 = \frac{\p}{\p t} + W^i \frac{\p}{\p x^i} 
\een
These two  differ by the wind term. The result of the boost is that
a magnetic vector potential  in one frame may appear as an electric vector
potential in the other.  

In fact, we shall show explicitly  later, that,
at least in the case of a axial symmetry, 
the effect of the wind is equivalent to an induced {\sl electric} connection
$A_\mu$. Thus, as we shall show in detail
later in the paper,  we could  in principle construct  a graphene analogue
of the {\sl  conformal geometry} of a rotating $2+1$ -dimensional black hole
such as that of  BTZ \cite{BTZ}. Actually as we shall show later, 
there is a fundamental  geometric obstacle, already encountered in \cite{Iorio}
to  obtaining a model system which extends all the way to the horizon.
   
   The paper is organized in the following way. In Section 2 we show that  only a part of the BTZ optical metric, which lies outside the outer horizon, can be mapped onto the  Beltrami Trumpet, an example of the  axisymmetric metrics which have an isometric   embedding into thee-dimensional Euclidean space  ${\Bbb E^3}$. In Section 3 we  embed the exact static  as well as  rotating BTZ  black hole optical metric into ${\Bbb E}^3$. In the latter case the metric is of the Zermelo type.   In Section 4 we derive the form of the external electro-magnetic field  applied to a graphene.  In Section 5 we analyze the Dirac equation for fermions confined to the graphed sheet  with  the applied elector-magnetic field; the   Zermelo axisymmetric metrics model an electric field, while  the Randers metrics model a  magnetic field in  a frame that is boosted relative to the Zerlemo metric frame. Conclusions  and possible other applications of  optical metrics are given in  Section 6.

\section{Surfaces of revolution} 

From now on we shall confine attention to axisymmetric metrics 
which may be isometrically embedded in Euclidean space ${\Bbb E}^3$ 
as  surfaces of revolution. We have
\ben
 h_{ij}dx^idx^j =d \rho ^2 + C^2(\rho) d \phi ^2 \,, \qquad 0\le \phi < 2 \pi \label{sr}
\,,\een
so that
\ben
C^2(\rho) = x^2 +y ^2 =R^2  
\een
and 
\ben
d\rho^2 = dR^2 + dz^2 \,.
\een
The Gauss curvature is given by
\ben
K= - \frac{C^{\prime \prime}}{C} \,.
\een

The example introduced  in \cite{Iorio} 
is given by  the {\it Beltrami Trumpet} 
\ben
C(\rho) = a \exp (-{\frac{\rho}{a}})  \,,\qquad \rho \ge 0,  
\qquad \Rightarrow  \qquad 
K=-\frac{1}{a^2} \,.
\een
If 
\ben
w= a \phi +i a \exp ({\frac{\rho}{a}})\,,
\een
then
\ben
 h_{ij} dx^i dx^j = \frac{ a^2  |dw|^2} { (\Im w) ^2 }\,,  
\een
which would be  the standard model of $H^2$ as the upper half complex plane,
if ${\Im  w} \ge 0$ but we must  quotient by the 
$\Bbb Z$ action  $w \rightarrow w+ 2 \pi a$
and moreover take  ${\Im  w} \ge a$ .
As is well known, the  Beltrami Trumpet is obtained by revolving 
a {\it tractrix} curve about the $z$-axis. As one may verify,
the tractrix  is the locus of one  end of a light rope of length 
$a$ to which is attached
to a heavy weight  as the other end is dragged along the negative $z$ axis.
Initially the rope occupies the interval $0\le x \le a$, $y=0\, z=0$
and there is a half cusp at $(a,0,0)$.

This failure to embed  all of $H^2/{\Bbb Z}$ is not an artifact of our
symmetry assumptions. A theorem of Hilbert implies
that we cannot get a non-singular  embedding  
of any complete metric of constant  negative curvature into three dimensional
Euclidean space. As we we see shortly this in turn implies that
no non-singular embedding of the optical metric of a $2+1$ dimensional
black hole can never reach the horizon. It can only extend a finite
way to the horizon, which is at infinite optical distance..

For vanishing angular momentum ($J=0$) 
the BTZ metric  (\ref{BTZm}) takes the form:
\ben
ds_{BTZ}  ^2 = -(\frac{ r^2}{l^2}-M) dt ^2  + \frac{d  r^2}
{(\frac{ r^2}{l^2}-M)  } +  r^2 d\phi ^2 \,, 
\qquad 0\le \phi < 2 \pi \,. 
\een
The associated {\it optical metric}  is
\ben
ds ^2_o = -dt ^2 +  \frac{d  r^2}
{(\frac{ r^2}{l^2}-M)^2   }  + \frac{ r^2}{ 
(\frac{ r^2}{l^2}-M)} d \phi ^2  \,, 
\qquad \Omega ^2 = (\frac{ r^2}{l^2}-M) \,.   \label{BTZso}
\een
We set 
\ben r_+= l \sqrt{M}\,, \qquad  a=2l \,,\qquad ( r-  r_+)
= \frac{1}{8}  r _+  \exp (- 2 \frac{\rho}{a} ) \,, 
\een
and expand about the horizon to see that 

\medskip \noindent 
{\it To lowest order the region
$\tilde r \ge \frac{9}{8}  \tilde r_+ $ of the optical metric of 
BTZ black hole maps onto
the  Beltrami spacetime.}

\section{Exact Static and Rotating BTZ Black Holes}

In this Section we shall  embed the exact BTZ optical geometry into ${\Bbb E} ^3$. 
We do that first for the  static  BTZ optical geometry (\ref{BTZso}).
Thus
\ben
{ ds^2_o } = -d{ t}^2 + l^4 \frac{dr^2} {(r^2-a^2) ^2} 
+ \frac{L^2r^2} {r^2-a^2} d \phi ^2\,,   
\een
where $a=l\sqrt{M}$. We introduce:
\ben C^2=  \frac{l^2r^2}{r^2-a^2} \,,
\een
and thus
\ben
dz^2 + dC^2 = \frac{l^4  dr^2}{(r^2-a^2) ^2} \,,\qquad \Rightarrow \qquad
dz^2 =l^2 \frac{(r^2-a^2 (l^2 +a^2 ) )\, dr ^2 }{(r^2-a^2 )^3 } 
\,.\een
Thus
\ben
\bigl ( \frac{dz}{dC} \bigr ) ^2 = \frac{1 + \frac{l^2}{a^2} -\frac{C^2}{l^2}} {C^2-l^2}  
\,.
\een
Clearly the embedding must stop at the radius for which
 $ C=\sqrt{l^2  + a^2   }$.
This is outside the horizon for which $C^2 \rightarrow \infty$.
The radial optical  distance $\rho$  is given by  
\ben
d \rho =  l^2 \frac{dr}{r^2-a^2}\,, 
\een
Thus
\ben
\frac{r}{a} =\coth(\frac{a}{l} \rho) \,,\qquad C= l\cosh (\frac{a}{l} \rho)  
\,,\een
and so the optical metric is 
\ben
ds_o^2 = -d{ t}^2 + d \rho ^2 + l^2 \cosh^2 (\frac{a}{l} \rho)  d\phi^2
\,,\een
and the BTZ metric itself is 
\ben
ds _{BTZ} ^2 = \frac{a^2}{ \sinh^2(\frac{a}{l}\rho) }
\Bigl  \{  -dt^2 + d \rho ^2 + l^2 \cosh^2 (\frac{a}{l} \rho)  d \phi ^2  \Big\}                  \,.\een
{\sl Note that the Gaussian curvature of the
spatial part of the BTZ optical metric
is of constant negative curvature}.  {\sl Note also that the BTZ optical geometry is locally of the
form ${\Bbb R} \times H^2$ and hence conformally flat.}
  
Assuming that  such graphene sheets  with negative constant curvature  
can be made in the Laboratory such BTZ Beltrami Trumpets could also be made.

We now turn to the  rotating  BTZ black hole which  has non-zero angular momentum $J$. Its  metric (\ref{BTZm})  is conformal to the Zermelo metric (\ref{Zermelo}) 
\ben
ds^2  = -dt ^2 + \frac{dr^2}{\Delta^2} + \frac{r^2} {\Delta} 
\bigl ( d \phi  -\frac{J}{2 r^2} dt \bigr )^2   
\, .\label{BTZz}
\een
and  the conformal factor $V=\Delta$, specified by (\ref{cf}).
The Zermelo metric  coefficients are thus of the form:
\bea
h_{ij}dx^i dx^j &=& \frac{dr^2}{\Delta ^2 } + \frac{r^2} {\Delta}  d \phi ^2  
\,,\nonumber\\
W^i \p_i&=& \frac{J}{2 r^2} \p_\phi \,.  \label{BTZzz}
\eea
In the near horizon limit  described above $h_{ij}$ is of  Beltrami Trumpet
form. 
  
\section{Gauge Fields} 

In this Section we study external elect-magnetic fields
applied to a graphene surface of revolution. 
Such external elector-magnetic fields  will induce a non
trivial ones on the surface of the graphene.

In the magnetic case, we are interested in two cases.
One in which the magnetic field in the bulk has constant magnitude$B_0$.
$B_0$ . The second is when the magnetic field normal to the surface 
$B_n$ 
has constant  magnitude $B_c$.  Suppose that
the bulk gauge field  $A$ and its field-strength $F$ are  given by  
\ben
A_\mu dx^\mu = A(C) d \phi \,, \qquad \Rightarrow \qquad F= \frac{A^\prime}{C} 
C d C  \wedge d \phi = \frac{A^\prime}{C} dx \wedge dy  
\,.\een 
In the embedding space this will correspond to a magnetic field
\ben
B_z = \frac{A^\prime}{C}  
\een
Thus an applied uniform  magnetic field  with strength $B_0$ has the gauge potential has $A(C)=B_0 C$. 
On the surface however since 
\ben
F= \frac{A^\prime(C) }{C}  \frac{d C}{d \rho}   d \rho  \wedge C  d \phi =
\frac{A^\prime(C) }{C}  \frac{d C}{d \rho}  \eta \,,\een 
where $\eta =d \rho  \wedge C  d \phi   $ is the area 2-form on the surface, 
the normal field strength $B_n$ is
\ben
B_n= \frac{A^\prime(C) }{C}  \frac{d C}{d \rho}  
\,.\een
Thus a constant field $B_0$  will produce a normal field
\ben
B_n=B_0  \frac{1}{C} \frac{d C}{d \rho} = 
B_0  \frac{a}{l}  \tanh ( \frac{a}{l} \rho)  
\een 
In order to produce a uniform field on the surface we need to set
\ben
\frac{A^\prime(C) }{C}  \frac{d C}{d \rho} = B_c= {\rm constant} \,. 
\een
Thus
\ben
\frac{dA}{d\rho} = B_c  C  \,.
\een
That is 
\ben
B_z= B_c ( \frac{dC}{d \rho} ) ^{-1} = \frac{lB_c}{a \sinh (\frac{a}{l} \rho) }  \,.
\een
Note that the angle $\psi$ that  the meridians make with the vertical 
direction is given by
\ben
\sin \psi = \frac{dC}{d \rho} \,, 
\een  
These formulae are geometrically rather obvious.
in particular we have 
\ben
B_n= B_z \sin \psi\, . 
\een 

An analogous analysis can also be performed for the electric field.
In this case we have a bulk vector potential
\ben
A_\mu\, dx^\mu= \Phi(R) dt\, ,  \label{elec}
\een
which would, if the graphene sheet were non-conducting,
induce a radial  electric field on the surface
\ben
F= \frac{d\Phi(\rho)}{d \rho} d \rho \wedge dt\,,  
\een
where $R=C(\rho)$. Thus
\ben
F= \frac{d\Phi(R)}{d R} \frac{d C}{d \rho}   d \rho \wedge dt\,.
\een

\section{The Dirac equation}
In this Section we study the Dirac equation for fermions confined to the graphene sheet.  The Dirac equation  is of the form:
\ben
\bigl( \gamma ^i \nabla _i + \gamma ^0 \p_t) \Psi =0\,,   \label{de}\ \ 
\een
where $(i,j)$  are dyad indices and 
\ben
d e^i= -\omega ^i\,_j \wedge e^j\, ,
\een
and
\ben
\nabla \Psi = d\psi  + \frac{1}{4} \omega_{ij} \gamma ^i \gamma ^j \Psi  \,. \een
For a surface of revolution metric (\ref{sr}) the pseudo-orthonormal one forms are (We use $-++$  signature.):
\ben
e^1= d \rho\,, \qquad e^2 = C(\rho) d \phi\,,
\een
and
\ben
d e^2= \frac{C^\prime}{C} e^1 \wedge e^2 \,,\qquad \Rightarrow  \qquad \omega_{21} =
  \frac{C^\prime}{C} e^2  \,.
\een
Thus
\ben
\nabla =d +  \gamma ^2 \gamma ^1 \frac{C^\prime}{2C} e^2\,,
\een
and  the Dirac equation (\ref{de}) can be cast in the form:
%\ben
%\bigl (\gamma ^1 \p_\rho + \gamma^2  \frac{1}{C}\p _\phi + 
 %\gamma ^2 \gamma ^2 \gamma ^1  \frac{C^\prime}{2C}                                + \gamma ^0 \p_t       \bigr ) \Psi =0                                           %\,. 
%\een
%That is 
\ben \gamma ^1 \frac{1}{\sqrt{C} }  \p_\rho( \sqrt{C} \Psi    )
 + \gamma^2  \frac{1}{C} \p _\phi \Psi                                  + \gamma ^0 \p_t  \Psi =0                                            
\,,\een
If we  set
\ben
\sqrt{C} \Psi = \tilde \Psi \,, \tilde \gamma ^i = \gamma ^0 \gamma ^i
\een
we  obtain \ben \tilde \gamma ^1  \p_\rho  \tilde \Psi    
 + \tilde \gamma^2  \frac{1}{C} \p _\phi \tilde  \Psi                                  - \p_t  \tilde \Psi =0 \label{reduced}                                    \,.        
\een
Both  $\gamma ^3$ and $\gamma ^0 \tilde \gamma ^1 \tilde \gamma ^2$  
commute with (\ref{reduced}) and themselves, and 
 have  eigenvalues $\pm$. The   solution can therefore be cast in terms of eigenvalues of these matrices.

We  now turn to the   axisymmetric optical Zermelo  metric, written in the form:
\ben
ds ^2 = - dt ^2 + d\rho ^2 + C^2(\rho) ( d \phi -W(\rho) dt ) ^2  
\, .
\een
Note that the rotating BTZ solution (\ref{BTZz}) is of this form.  A pseudo-orthonormal   basis of one-forms is defined as:
\ben
e^0=dt\,,\qquad e^1 = d \rho \,,\qquad e^2 = C(d \phi - W dt) 
\,,\een
and a dual basis of vector fields by
\ben
e_0= \p _t + W \p_\phi \,,\qquad e_1 =\p _\rho\,,\qquad e_2 = \frac{1}{C} \p _\phi \,. 
\een
Note that $e_a \phi=  \delta ^0_a W(\rho)  $ and so the Dreibein $e_a$  
is differentially rotating. Using the formulae
\ben
d e^a = - \omega ^a\,_b \wedge e^b \,, \qquad  \omega_{ab} = \eta_{ac} \omega ^c\,_b = - \omega _{ba} \,,  
\een
with $a=0,1,2$ and $\eta_{ab}= {\rm diag} (-1,1,1)$ 
we find 
%{\bf should be checked!}- I have rechecked it explicitly!
\ben
\omega_{01}= -\half C W^\prime e^2\,,\qquad 
\omega_{02}= -\half C W^\prime e^1 \,,\qquad 
\omega_{21}= -\half C W^\prime e^0 + \frac{C^\prime}{C} e^2  
\,.\een
The massless Dirac equation may be written
as
\ben
\Bigl ( \gamma ^a e_a + \frac{1}{4} 
\gamma ^a \omega_{abc} \gamma ^b \gamma ^c
\Bigr ) \Psi =0\,,   
\een
where $\omega_{abc} $ are what are  sometimes 
called {\it Ricci rotation coefficients} 
\ben
\omega _{bc} =   e^a  \omega _{abc}  \,.
\een
Thus 
\ben
\Bigl (  \gamma ^ 1 \bigl( \p_\rho  + \frac{1}{2}\frac{C^\prime}{C}  \bigr ) +  
\gamma ^2 \frac{1}{C} \p_\phi 
+ \gamma ^0 (\p_t + W  \p_\phi)     
+\frac{1}{4} \gamma ^0 \gamma ^1 \gamma ^2  C W^\prime   \Bigr )  \Psi =0   
\,.\een
If we choose 
\ben
\gamma ^0= i\sigma_2\,,\qquad \gamma ^1= \sigma _1\,,\qquad \gamma ^2= \sigma_3
\,.\een
then $\gamma^0\gamma ^1 \gamma ^2 =1$ and we get a position dependent
``mass-like ``  term and a connection term. 
If $\Psi \propto e^{-i\omega t + i m \phi} $ we have that 
\ben
-ie A_0 = im  W \,,\qquad \Rightarrow eA_0= - m W
\,,\een
and the effective {\sl electric}  field in what is actually a 
rotating  frame is
\ben
eF= edA =m W^{\prime } d \rho \wedge d t =
mW^{\prime } e^1  \wedge  e^0 \,.  
\een
Therefore a stationary Zermelo metric  induces in the Dirac equation an effective,  position dependent radial  electric field.
The bulk electric potential  $A_\mu\, dx^\mu=\Phi(R) dt$ (\ref{elec})  is obtained by setting
\ben
\Phi(R)= \frac{m}{e} W(\rho) \,.
\een
From (\ref{BTZm}) and (\ref{cf}) we have  
\ben
R^2= \frac{r^2}{\frac{r^2}{l^2} -M + \frac{J^2}{4r^2} }
\een
and 
\ben
W= \frac{J}{2r^2}\,, \qquad  \Rightarrow \qquad r^2= \frac{J}{2W}\,.     
\een
Elimination leads to the quadratic equation
\ben
  \frac{1}{l^2} -\frac{1}{R^2}   -2 \frac{WM}{J} + W^2 =0 \,,  
\een
which may be solved to give $\Phi(R)$.

We could have  solved the Dirac equation in the   Randers 
form of the metric (\ref{Randers}). We define
\ben 
e^0=(dt + \omega d\phi) \,,\qquad e^\rho = d \rho\,, \qquad e^ \phi = C(\rho) d \phi \,.
 \een
\ben
e_0 = \frac{\p}{\p t} \,,\qquad e_\rho  =  \frac{\p}{\p \rho } \,,\qquad 
e_\phi= \frac{1}{C}
\bigl (\frac{\p}{\p \phi} - \omega  \frac{\p}{\p t} \bigl )  \,. 
\een
We find 
\ben
\omega ^\phi\,_\rho= -\frac{\omega ^\prime}{2C} e^0 + \frac{C^\prime}{C} e^\phi\, ,  \qquad
\omega ^0\,_\rho =  \frac{\omega ^\prime}{2C} e^\phi\, ,  \qquad
\omega ^0\,_\phi =- \frac{\omega ^\prime}{2C} e^\rho \, .
\een
Because in this case the roles of $t$ and $\phi$ have essentially  been  interchanged,
we now find that there is an effective {\sl magnetic} vector potential
in the Dirac equation. Therefore, as pointed out in the introduction the magnetic vector potential in the Randers frame appears as an electric potential in the  Zermelo one.

\section{Conclusions}

We argue that the curved graphene sheet with negative constant curvature in  the externally applied magnetic field  could be modeled by considering a stationary optical metric of the  Zermelo  form which is conformal to the BTZ black hole.  In particular,  the electric  field on the surface of the graphene can be modeled with  the {\it wind} of the  Zermelo metric.  On the other hand the Randers metric models a magnetic field, related to the electric one of the Zermelo frame by a boost. 
Furthermore, we establish that there is  a fundamental geometric obstruction to obtain a model that extends all the  way to the BTZ black hole horizon. 
We   model the low energy electron excitations  of such graphene sheets by  studying solutions of the $2+1$-dimensional 
massless Dirac equation in the  stationary optical metric, conformal to the BTZ black hole.

Related analyses can be applied to any other system described with a 
two-dimensional curved surface $\Sigma$.  We should also  point out that there are  other possible embeddings of a surface  with 
constant negative curvature  which are  not a surface of revolution but a 
twisted   Beltrami Trumpet. It embeds  more of $H^2$ than the  Beltrami Trumpet. It would  be interesting to further study implications of such more general embeddings. 

\vskip 0.5cm
\noindent {\bf Acknowledgments:} \ \ 
\noindent
 G.W.G thanks the UPenn Department of Physics \& Astronomy Department  for hospitality. 
MC is supported by the DoE 
Grant DOE-EY-76-02- 3071, the NSF RTG DMS Grant 0636606, the Fay R. 
and Eugene L. Langberg Endowed Chair  and the Slovenian
Research Agency (ARRS).

\end{document}